\documentclass[final,3p,times,twocolumn]{elsarticle}

\usepackage{epsfig}
\usepackage{amssymb}
\usepackage{amsmath}
\usepackage{dsfont}

\usepackage{graphicx,hyperref}
\usepackage{bm}
\usepackage{subfigure}

\newcommand{\beq}{\begin{equation}}
\newcommand{\eeq}{\end{equation}}
\newcommand{\bea}{\begin{eqnarray}}
\newcommand{\eea}{\end{eqnarray}}
\newcommand{\nn}{\nonumber \\}

\newcommand\eqn[1]{(\ref{#1})}      
\newcommand\Eqn[1]{Eq.~(\ref{#1})}  
\newcommand\Fig[1]{Fig.~\ref{#1}}  

\newcommand{\tr}{\hbox{tr}}
\newcommand{\bartheta}{\,\bar{\!\theta}}
\newcommand{\thetab}{{\bartheta}}
\newcommand{\cb}{{\bar c}}

\journal{Physics Letters B}

\begin{document}

\begin{frontmatter}



\title{Lifting the Gribov ambiguity in Yang-Mills theories}


\author{J. Serreau}

\address{APC, AstroParticule et Cosmologie, Universit\'e Paris Diderot, CNRS/IN2P3, CEA/Irfu, Observatoire de Paris, Sorbonne Paris Cit\'e,\\ 10, rue Alice Domon et L\'eonie Duquet, 75205 Paris Cedex 13, France}
\author{M. Tissier}
\address{LPTMC, Laboratoire de Physique Th\'eorique de la Mati\`ere Condens\'ee, CNRS UMR 7600, Universit\'e Pierre et Marie Curie, \\ boite 121, 4 pl. Jussieu, 75252 Paris Cedex 05, France}

\begin{abstract}
  We propose a new one-parameter family of Landau gauges for
  Yang-Mills theories which can be formulated by means of
  functional integral methods and are thus well suited for analytic
  calculations, but which are free of Gribov ambiguities and avoid the
  Neuberger zero problem of the standard Faddeev-Popov
  construction. The resulting gauge-fixed theory is perturbatively
  renormalizable in four dimensions and, for what concerns the calculation of ghost and gauge field
  correlators, it reduces to a massive
  extension of the Faddeev-Popov action. We study the renormalization
  group flow of this theory at one-loop and show that it has no Landau
  pole in the infrared for some -- including physically relevant --
  range of values of the renormalized parameters.
\end{abstract}

\begin{keyword} Yang-Mills Theory \sep gauge-fixing \sep Gribov ambiguities \sep infrared correlation functions


\end{keyword}

\end{frontmatter}



\section{Introduction}\label{sec:introduction}
In his seminal work on non-abelian gauge theories \cite{Gribov77}, Gribov showed that
the issue of fixing a gauge is far more complicated than in the
abelian case and may be hampered by some ambiguities. For instance,  for SU($N$) Yang-Mills (YM) theories,
the Landau gauge condition\footnote{We consider the Euclidean
  theory in $d$ dimensions. When necessary, a lattice discretization is implicitly considered.}  $\partial_\mu
A_\mu=0$ is satisfied by many field
configurations that are equivalent up to gauge transformations. To
completely fix the gauge one has to specify how to deal with these
Gribov copies.

For Landau gauges, Gribov copies are given by the extrema of the
functional $F[A,U]=\int d^dx\, \tr(A^U)^2$ with respect to $U$ for a
given field configuration $A$, where $A^U$ is the transform of
$A$ under the gauge transformation $U$. The
standard Faddeev-Popov (FP) procedure corresponds to summing over all
extrema of $F[A,U]$ (minima, maxima and saddle points) with a minus
sign for the Gribov copies whose FP operator has an odd number of
unstable directions \cite{Gribov77}. For compact gauge groups, such as SU($N$), this
alternation of sign leads to an exact compensation of contributions
from the various Gribov copies for gauge invariant quantities. As a consequence, formal expressions
of physical observables in fact appear as an undetermined $0/0$
quotient \cite{Neuberger:1986vv}. This so-called
Neuberger zero problem forbids the use of e.g. lattice regularizations
to devise a nonperturbative version of the gauge-fixed theory. It is
usually disregarded in the high-energy, perturbative regime, where Gribov copies are thought to be harmless and are simply
ignored. Another drawback of the FP gauge-fixing is that the resulting perturbation theory shows a Landau pole in
the infrared (IR) -- where the running coupling constant diverges -- and is
thus useless for studying the low energy properties of the theory.

Several other Landau gauge-fixing schemes can be considered. One can
for instance pick up a particular minimum of $F[A,U]$ for each field
configuration $A$. This has the advantage of being relatively easy to
implement for numerical calculations on the lattice \cite{Boucaud:2011ug}.  However,
depending on the algorithm used to find the minimum, different Gribov
copies will be attained, corresponding to different choices of
gauges. This renders the comparison between different calculations (of non gauge invariant quantities)
somewhat tricky. In particular, it has been shown that different
choices of algorithms may lead to a variation of the ghost propagator
by up to 10$\%$ at low momentum  \cite{Sternbeck:2005tk}.
  
One way to cope with this issue is to choose the
absolute minimum of $F[A,U]$ for each $A$, the so-called absolute Landau
gauge \cite{Dell'Antonio:1991xt}. This is clearly a pretty hard numerical task since this functional typically presents numerous local minima. Gribov and Zwanziger \cite{Gribov77,Zwanziger89} proposed to restrict the functional integration over the $A$ field to the first Gribov region (corresponding to the minima of $F[A,U]$). However, this region is not free of Gribov ambiguity \cite{Vandersickel:2012tz}.
Recently, a refined version of this Gribov-Zwanziger approach was
shown to lead to predictions for the gauge field correlation functions
that agree well with lattice simulations \cite{Dudal08}.

In this letter, we propose a new one-parameter family of Landau gauges
which are free of Gribov ambiguities and which turn out to be 
simple to implement for practical perturbative calculations. This is based on taking a particular
average over Gribov copies of each gauge field configuration. These are good
gauge-fixings in the sense that gauge-invariant quantities are independent of the gauge fixing procedure. We also point
out that this family of gauges interpolates between the FP and absolute Landau gauges. 
A similar averaging procedure, however not restricted to Gribov copies in the Landau gauge, was proposed in \cite{Parrinello:1990pm}. Another one-parameter family of Landau gauges was proposed in \cite{Maas:2009se}.

One important aspect of the gauge fixing procedure proposed here is
that it can be fully implemented by means of standard functional integral techniques and is thus well suited for analytic
calculations.  Indeed, our procedure amounts to taking a particular
average over Gribov copies of each gauge field configuration. This
average can be dealt with by means of a standard replica trick --
borrowed from the theory of disordered systems in condensed matter
physics -- which, together with a Faddeev-Popov (FP) construction, allows for
a simple path integral formulation. Another key aspect is that our
procedure does not suffer from the Neuberger zero problem of the
standard FP construction. It is in this sense a fully
justified field-theoretical description of YM theory which
takes into account the Gribov ambiguity from first principles \cite{Quadri:2010vt}. 

We show that the model is perturbatively
renormalizable in four dimensions thanks to the underlying symmetries. We also demonstrate that, within perturbation theory, correlation functions 
of YM fields are identical to those obtained in a massive
extension of the usual FP gauge-fixed action, which is a particular case of 
the Curcci-Ferrari (CF) model \cite{Curci76}. This provides a solid
first-principle field theoretical justification for the
phenomenological approach recently proposed in
\cite{Tissier_10,Tissier_11}. One-loop calculations in this model were shown to reproduce
quantitatively lattice results for the gluon and ghost propagators in the
Landau gauge. Finally, we discuss the renormalization group flow of the present gauge-fixed YM theory at one-loop 
and show that it is free of Landau pole for some range of renormalized parameters.

\section{The gauge-fixing procedure}

The main idea of this work is to deal with Gribov copies not by
choosing a unique one as is often done, but by averaging over them
with a given weight. To compute usual YM
correlators, we first average over Gribov copies for each given
gauge field configuration $A$
and then perform an average over the gauge field configurations with
the usual YM weight.

To be specific, we consider a SU($N$) YM theory in the Landau
gauge. The classical action reads 
\beq S_{\rm
  YM}[A]=\frac{1}{2}\int_x\tr \left(F_{\mu\nu}\right)^2\,, 
\eeq 
where $F_{\mu\nu}=\partial_\mu A_\nu-\partial_\nu
A_\mu -ig_0[A_\mu,A_\nu]$, $\int_x\equiv\int d^dx$ and $g_0$
is the (bare) coupling constant. Here and below, when a field is written without color index, a
contraction with the generators of the group in the appropriate
representation is understood.
Our convention for the generators is $ \tr \,T^a T^b=\frac 12 \delta^{ab}$ and $[T^a,T^b]=if^{abc}T^c$.

For any operator $\mathcal{O}[A]$, we define the average over the Gribov copies of a given field configuration $A$ as:
\begin{equation}
  \label{eq_average_G}
  \langle\mathcal O[A]\rangle=\frac{\sum_i \mathcal O [A^{U_i}]s(i)e^{-S_{\rm \!W}[A^{U_i}]}}{\sum_i s(i)e^{-S_{\rm \!W}[A^{U_i}]}}
\end{equation}
with weight factor
\begin{equation}
  \label{eq_z}
S_{\rm \!W}[A]=  \beta_0\int_x\,\tr\left(A_\mu\right)^2\,.
\end{equation}
and where the gauge transform $A^U$ is defined as: 
\begin{equation}
  \label{eq_gauge_transfo}
  A_\mu^U=UA_\mu U^{-1}+\frac i {g_0}U\partial_\mu U^{-1}\,.
\end{equation}
In \Eqn{eq_average_G}, the sum runs over all Gribov copies, that is over all $U_i$ which
satisfy $\partial_\mu A_\mu^{U_i}=0$ or, equivalently, over all extrema
of $S_{\rm \!W}[A^{U_i}]$, for a given $A$; $s(i)$ is the sign of the
functional determinant of the FP operator 
\beq
 \left(\delta^{ab}\partial^2-g_0f^{abc}A_\mu^c\partial_\mu\right)\delta^{(d)}(x-y)
\eeq
taken at $A=A^{U_i}$. Finally,
$\beta_0$ is a free parameter. The averaging procedure \eqn{eq_average_G}-\eqn{eq_z} is inspired from the one proposed in \cite{Tissier:2011zz} in the context of the Random Field Ising Model. Note that a somewhat similar gauge-fixing has been proposed in \cite{Parrinello:1990pm} where, however, the average was not restricted to Gribov copies in the Landau gauge. This difference is essential e.g. in making the present proposal renormalizable, as we discuss below. Note also that a similar average over Gribov
copies was considered in \cite{Bogolubsky:2007bw} to implement a
simulated annealing as an efficient method to approach the
absolute Landau gauge. Here, however, the sum runs over all Gribov
copies, not only on those in the first Gribov region. Notice though that, for $\beta_0$ not too small, the copies outside the first region are suppressed by the weight factor \eqn{eq_z}. 

Actually, the limit $\beta_0\to \infty$ selects the absolute minimum $U=U_{\rm abs}$ of
$S_{\rm \!W}[A^U]$ and our averaging procedure simply corresponds to the
absolute Landau gauge:
\begin{equation}
  \label{eq_absolute}
  \lim_{\beta_0\to \infty}\langle\mathcal O[A]\rangle=\mathcal O[A^{U_{{\rm abs}}}]\,.
\end{equation}
In the opposite limit $\beta_0\to0$, all Gribov copies
contribute the same in the average \eqn{eq_average_G} up to the sign
factor $s(i)$. Since there are as many contribution with each sign,
the denominator in \eqn{eq_average_G} vanishes: $\sum_is(i)=0$. For any $\beta_0>0$, the degeneracy is lifted
and the denominator in \eqn{eq_average_G} differs from zero in
general,\footnote{Strictly speaking, we cannot exclude that for a
  given value of $\beta_0$ some field configurations yield a vanishing
  denominator. We assume such field configurations to be of zero
  measure.}
 which solves the Neuberger zero problem.  

Once the average \eqn{eq_average_G} over Gribov copies has been performed for each individual field configuration we perform the usual average over YM field configurations, hereafter denoted by an overall bar:
\begin{equation}
  \label{eq_av_A}
\overline {\mathcal O[A]}=\frac{\int\mathcal
  DA\, \mathcal O[A]e^{-S_{\rm YM}[A]}}{\int\mathcal
  DA\, e^{-S_{\rm YM}[A]}}
\end{equation}
To summarize, computing a given observable ${\cal O}$ in our gauge fixing procedure amounts to first average over Gribov copies and then to average over YM field configurations, that is
\begin{equation}
  \label{eq_avav}
  \overline{\langle\mathcal O[A]\rangle}\,.
\end{equation}

A crucial remark is in order here: observe that gauge-invariant operators such that $\mathcal{O}_{\rm inv}[A^U]={\cal O}_{\rm inv}[A]$, are blind to the average \eqn{eq_average_G}:
$ \langle\mathcal O_{\rm inv}[A]\rangle=\mathcal O_{\rm inv} [A]$,
which guarantees that our gauge fixing procedure does not affect physical observables. In particular, one has 
\beq
 \overline{\langle{\cal O}_{\rm inv}[A]\rangle}=\overline{{\cal O}_{\rm inv}[A]}\,.
\eeq 
It is crucial to introduce the denominator in \eqn{eq_average_G} in order for this fundamental property to hold. 

\section{Functional integral formulation}

The notations used in the previous section where chosen to make
contact with the physics of disordered systems \cite{young}, the theoretical description of which also typically involves a two step averaging procedure: One has to average first over thermal fluctuations for a given realization of the (quenched) disorder and then over the disorder. In the above proposal for YM theories, the analog of the first (thermal) average is the one over Gribov copies, \Eqn{eq_average_G}, for fixed gauge field $A$. The latter thus plays the role of the disorder field to be averaged over in the second step. Although such two-step averages are rather easily implemented in numerical simulations, they are technically challenging for analytical calculations. The method of replica is designed to handle this issue. We show here how to adapt it to the problem at hand.

Let us first rewrite the average over Gribov copies as a functional integral. Following the standard FP procedure, the sums over Gribov copies appearing in \eqn{eq_average_G} -- including the sign $s(i)$ -- can be represented by a functional integral over a SU($N$) matrix field $U$, FP ghost (Grassmann) fields $c$ and $\bar c$ and a Lagrange multiplier $h$. One has, in particular,
\begin{equation}
  \label{eq_gribov_FP}
  \sum_i \mathcal F [A^{U_i}]s(i)=\int {\cal D}(U,c,\bar c,h)\,\mathcal F [A^U] \,e^{-S_{{\rm FP}}[A^U\!,c,\bar c,h]}    
\end{equation}
where ${\cal D}(U,c,\bar c,h)\equiv {\cal D}U{\cal D}c{\cal D}\bar c{\cal D}h$, with ${\cal D}U$ the Haar measure on the gauge group, and
\begin{equation}
  \label{eq_fp}
  S_{{\rm FP}}[A,c,\bar c,h]\!=\!\!\!\int_x\Big\{\partial_\mu\bar
    c^a(\partial_\mu c^a+g_0f^{abc}\!A_\mu^b c^c)+ih^a\partial_\mu A_\mu^a\Big\} 
\end{equation}
is the usual FP action. Collecting the set of FP fields $U$, $c$, $\bar c$ and $h$ in a single symbol ${\cal V}$ -- we shall see shortly how this can be realized explicitly in a supersymmetric formulation --, the average \eqn{eq_average_G} over Gribov copies can be written as
\beq
\label{eq:fact1}
 \langle {\cal O}[A]\rangle=\frac{\int{\cal D}{\cal V}\,{\cal O}[A^U]\,e^{-S_{\rm \!GF}[A,{\cal V}]}}{\int{\cal D}{\cal V}\,e^{-S_{\rm \!GF}[A,{\cal V}]}}\,,
\eeq
with the gauge fixing action
\beq
\label{eq:gf}
 S_{\rm \!GF}[A,{\cal V}]=S_{\rm \!W}[A^U]+S_{\rm FP}[A^U\!,c,\bar c,h]\,.
\eeq

The denominator of the previous expression can be formally rewritten using the replica trick \cite{young}
\begin{equation}
  \label{eq_replica1}
  \frac{1}{\int \mathcal D {\cal V}
    \, e^{-S_{\rm \!GF}[A,{\cal V}]}}=\lim_{n\to0}\int\prod_{k=1}^{n-1}\left( \mathcal D {\cal V}_k
    \, e^{-S_{\rm \!GF}[A,{\cal V}_k]}\right)\,.
\end{equation}
Here, the limit is to be understood as the value of the (analytically continued) function of $n$ on the right-hand-side when $n\to0$.
The average over the disorder field $A$ can then be formally written as
\begin{equation}
  \label{eq_average2}
  \overline{\langle{\cal O}[A]\rangle}=\lim_{n\to 0}\frac{\int\mathcal D A\left(\prod_{k=1}^n \mathcal D {\cal V}_k\right)\,{\cal O}[A^{U_1}]\, e^{-S[A,\{{\cal V}\}]}}{\int{\cal D}A\,e^{-S_{\rm YM}[A]}}\,,
\end{equation}
where
\beq
\label{eq:action1}
 S[A,\{{\cal V}\}]=S_{\rm YM}[A]+\sum_{k=1}^{n}S_{\rm \!GF}[A,{\cal V}_k]\,.
\eeq
Finally, using \Eqn{eq_average2} with ${\cal O}[A]=1$, one can rewrite \eqn{eq_average2} in the more convenient form
\begin{equation}
  \label{eq_average2bis}
  \overline{\langle{\cal O}[A]\rangle}=\lim_{n\to 0}\frac{\int\mathcal D A\left(\prod_{k=1}^n \mathcal D {\cal V}_k\right)\,{\cal O}[A^{U_1}]\, e^{-S[A,\{{\cal V}\}]}}{\int\mathcal D A\left(\prod_{k=1}^n \mathcal D {\cal V}_k\right)\, e^{-S[A,\{{\cal V}\}]}}\,.
\end{equation}
Here, the choice of the replica $k=1$ is arbitrary because of the obvious symmetry between replicas.

It is useful, for perturbative calculations, to explicitly factor out the volume of the gauge
group $\int{\cal D}U$. This can be done e.g. by performing the change of variables 
$A\to A^{U_1}$ and $U_{k}\to U_k U_1^{-1}$,  $\forall\,\, k>1$ in \eqn{eq_average2bis}. Renaming $(c_1,\cb_1,h_1)\to (c,\cb,h)$, we get
\begin{equation}
  \label{eq_average2ter}
  \overline{\langle{\cal O}[A]\rangle}=\lim_{n\to 0}\frac{\int\mathcal D (A,c,\bar c,h,\{{\cal V}\})\,{\cal O}[A]\, e^{-S[A,c,\bar c,h,\{{\cal V}\}]}}{\int\mathcal D (A,c,\bar c,h,\{{\cal V}\})\, e^{-S[A,c,\bar c,h,\{{\cal V}\}]}}\,,
\end{equation}
with $\mathcal D (A,c,\bar c,h,\{{\cal V}\})\equiv\mathcal D (A,c,\bar c,h)\left(\prod_{k=2}^n \mathcal D {\cal V}_k\right)$ and
\begin{align}
  &S[A,c,\bar c,h,\{{\cal V}\}]=S_{{\rm YM}}[A]+S_{\rm \!W}[A]+S_{{\rm FP}}[A,c,\cb,h]\nn
  \label{eq_action2}
  &\qquad\quad+\sum_{k=2}^n \left(S_{\rm \!W}[A^{U_k}]+S_{{\rm FP}}[A^{U_k},c_k,\bar c_k,h_k]\right).
\end{align}

As mentioned previously, the present theory can be given an elegant supersymmetric formulation, which proves powerful. To this purpose, we specify the symbol ${\cal V}$ introduced before as the following matrix superfield:
\begin{equation}
  \label{eq_susy}
  \mathcal V(x,\theta,\bartheta)=\exp \left\{ig_0\left(\bartheta c+\bar
    c\theta+\bartheta\theta \tilde h\right)\right\}U
\end{equation}
living on a superspace made of the original $d$-dimensional Euclidean space ($x$) supplemented by two Grassmann dimensions ($\theta,\bartheta$), which we collectively denote by $\underline{\theta}$. Here, $\tilde h=ih-i{g_0\over2}\{\bar c,c\}$ and the $x$-dependence only appears through the fields $U$, $c$, $\bar c$ and $h$. The first factor on the right-hand-side of \eqn{eq_susy} is trivially an element of SU($N$) and so is the superfield $\mathcal V$.

Defining further the ``super gauge transform'' $A^{\cal V}$ as in (\ref{eq_gauge_transfo}): $A_\mu^{\cal V}={\cal V}A_\mu {\cal V}^{-1}+\frac i {g_0}{\cal V}\partial_\mu {\cal V}^{-1}$, it is straightforward to show that the averaging weight in
(\ref{eq_gribov_FP}) takes the remarkably simple form:
\begin{equation}
  \label{eq_susy2}
S_{\rm \!W}[A^{U}] +S_{{\rm FP}}[A^{U},c,\bar
c,h]=\int_{x,\underline{\theta}}\tr(A^{\mathcal V})^2      
\end{equation}
with the notation
\begin{equation}
  \label{eq_note_int_grass}
\int_{x,\underline{\theta}}=\int d^dxd\theta d\bartheta\,g^{1/2}(\theta,\thetab)  \,,
\end{equation}
where the factor $g^{1/2}(\theta,\thetab)=\beta_0\thetab\theta-1$ on the right-hand-side can be seen as the invariant measure associated with curved Grassman dimensions, as discussed in Ref. \cite{Tissier_08}.

To gain more insight about the structure of the present theory, we write, alternatively,
\beq
\label{eq:nonlin}
 \tr\left(A_\mu^{\mathcal V}\right)^2=\tr\left(A_\mu-{i\over g_0}{\cal V}^{-1}\partial_\mu{\cal V}\right)^2=-{1\over g_0^2}\tr \left({\cal V}^{-1}D_\mu{\cal V}\right)^2
\eeq
where we introduced the covariant derivative $D_\mu{\cal V}\equiv\partial_\mu{\cal V}+ig_0{\cal V}A_\mu$. We see that the action \eqn{eq:action1}, with \eqn{eq:gf} and \eqn{eq_susy2}-\eqn{eq:nonlin}, thus describes a collection of $n$ gauged supersymmetric nonlinear sigma models \cite{Henneaux:1998hq}. It is invariant under the super gauge transformation $A\to A^{\cal U}$ and ${\cal V}_k\to{\cal V}_k{\cal U}^{-1}$, $\forall k=1,\ldots,n$. 
This symmetry gets explicitly broken after one replica is singled out to extract the volume of the gauge group. The action \eqn{eq_action2} describes a gauged-fixed theory for $n-1$ gauged supersymmetric nonlinear sigma models with CF gauge fixing $S_{\rm \!W}[A]+S_{{\rm FP}}[A,c,\cb,h]$:
\begin{align}
\label{eq:action}
  S[A,c,\bar c,h,\{{\cal V}\}]&=S_{{\rm YM}}[A]+S_{\rm \!W}[A]+S_{{\rm FP}}[A,c,\cb,h]\nn
  &-{1\over g_0^2}\sum_{k=2}^n \int_{x,\underline{\theta}}\tr \left({\cal V}_k^{-1}D_\mu{\cal V}_k\right)^2.
\end{align}

\section{Renormalizability}

We now prove the perturbative renormalizability of the model \eqn{eq_action2}-\eqn{eq:action} in $d=4$. This is a rather non-trivial result given the presence of nonlinear sigma models, which are in general renormalizable in $d=2$. 
For simplicity, we focus on the SU($2$) case for which $T^a=\sigma^a/2$ with $\sigma^a$ the Pauli matrices and $f^{abc}=\epsilon^{abc}$ the L\'evy-Civita symbol. Our proof follows standard arguments \cite{WeinbergBook} and consists in identifying all local terms of mass dimension less or equal to four \footnote{This relies on Weinberg's theorem and assumes, in particular, that the free propagators decrease sufficiently fast at large momentum. We show in the next section that all free propagators behave as $1/p^2$ at large $p$, which is a sufficient condition.} in the effective action $\Gamma$ compatible with the symmetries of the theory. 

Let us first list the symmetries of the action \eqn{eq:action} that are realized linearly. Apart from the trivial global SU($2$) color symmetry and the isometries of the $\mathds{R}^4$ Euclidean space, one has a shift symmetry for the $\cb$ field ($\cb\to\cb +{\rm const.})$, the net ghost number conservation ($c\to e^{i\epsilon}c$, $\cb\to e^{-i\epsilon}
  \cb$) and the isometries of the curved Grassmann space. The latter only impact the superfields: $\mathcal V_k\to\mathcal
  V_k+X^{\underline{\theta}}\partial_{\underline{\theta}}\mathcal V_k$ where $X^{\underline{\theta}}$ is one of the five independent Killing vectors on the Grassmann space \cite{Tissier_08}. At the level of the effective action, these symmetries simply imply that terms involving Grassmann variables should be written in a covariant way: integrals always come with proper integration measure, see \eqn{eq_note_int_grass}, and derivatives are contracted with proper tensors \cite{Tissier_08}. An important remark to be made is that these transformations apply to each individual replica superfield $\mathcal V_k$, independently of the others. This implies that each such superfield always comes with its own set of Grassmann variables. There is also a discrete symmetry under permutation of the replicas: ${\cal V}_k\leftrightarrow{\cal V}_l, \,k,l=2,\ldots,n$. 
 
These transformations are also symmetries of the effective action $\Gamma$ and directly constrain the possible divergent terms. We also use the fact that the linear term involving the field $h$ is not renormalized: $\delta\Gamma/\delta h^a=\delta S/\delta h^a=i\partial_\mu A^a_\mu$, and that the choice of the replica $k=1$ singled out in \eqn{eq_average2ter}-\eqn{eq_action2} being arbitrary, the divergences associated with the fields $c$, $\cb$ and $h$ are the same as those associated with $c_k$, $\cb_k$ and $h_k$ for $k\ge2$.\footnote{For instance, upon the
change of variables $A\to A^{U_2}$, $U_k\to U_kU_2^{-1}$ for $k>2$, $U_2\to U_2^{-1}$ and $c\leftrightarrow c_2$, $\cb \leftrightarrow \cb_2$ and $h\leftrightarrow h_2$, one gets that it is now the replica $k=2$ which is singled out.}
  
The action \eqn{eq:action} also admits non-linear symmetries. One is a BRST-like symmetry, corresponding to the infinitesimal transformation:
\begin{equation}
  \label{eq_brst}
  \begin{split}
    sA_\mu^a&=\partial_\mu c^a+g_0\epsilon^{abc}A_\mu^bc^c\,,\qquad s\cb^a=ih^a\,,\\
sc^a&=-\frac {g_0}2f^{abc}c^bc^c\,,\qquad\qquad s(ih^a)=\beta_0 c^a\,,\\
s\mathcal V_k&=-ig_0\mathcal V_k c\,, \quad k=2,\ldots,n\,.
  \end{split}
\end{equation}
It is useful to employ the decomposition of SU($2$) matrices in terms of a unit 4-vector:
\begin{equation}
\label{eq:decomp}
  \mathcal V_k=n^0_k\mathds{1}+in^a_k\sigma^a\,,
\end{equation}
where $(n_k^0)^2+n_k^an_k^a=1$. The BRST transformation of the constrained superfield ${\cal V}_k$ then reads:
\begin{equation}
  \label{eq_brst2}
    sn^0_k=\frac {g_0}{2}n_k^ac^a\,\,\,\,{\rm and}\,\,\,\,
   sn^a_k=\frac {g_0}2\left(-n_k^0 c^a+\epsilon^{abc}n_k^bc^c\right)\,.
\end{equation}
Note that, just as $\mathcal V_k$, $n_k^0$ and $n_k^a$ are superfields, that is functions of $(x,\underline{\theta})$. In the following we choose $n_k^a$ to be the unconstrained superfields and $n_k^0=(1-n_k^an_k^a)^{1/2}$. 

The action \eqn{eq:action} is also invariant under ${\cal V}_k\to {\cal U}{\cal V}_k$, where the SU(2) matrix ${\cal U}={\cal U}(\underline{\theta})$ can be local in Grassmann variables. This symmetry is nonlinear since ${\cal V}_k$ is a constrained superfield. Each replica superfield can be transformed independently of the others and there are thus $3\times (n-1)$ such symmetries. The infinitesimal transformations read $\delta_k^a\mathcal V_l=i\delta_{kl}\sigma^a\mathcal V_l$ or, in terms of the decomposition \eqn{eq:decomp},
\begin{equation}
  \label{eq_nl_color2}
  \delta_k^an^0_l=-\delta_{kl}n^a_l\,\,\,\,{\rm and}\,\,\,\,
  \delta_k^an^b_l=\delta_{kl}\left(\delta^{ab}n^0_l+\epsilon^{abc}n_l^c\right)\,.
\end{equation}

To derive Slavnov-Taylor identities associated with the nonlinear symmetries \eqn{eq_brst} and \eqn{eq_nl_color2}, one introduces (super)sources coupled to both the (super)fields and their variations under $s$ and $\delta^a_k$. We define:\footnote{Note that the symmetries $s$ and $\delta^a_k$ are not nilpotent. However, the transformations $s^2$, $s\delta^a_k=\delta^a_k s$ and $\delta^a_k\delta^b_l$ can be fully expressed in terms of the (super)fields at hand and their variations under either $s$ or $\delta^a_k$. Therefore they do not require independent (super)sources.}
\begin{align}
  \label{eq_sources}
  S_1&=\int_x\left\{J_\mu^a A_\mu^a\!+\!\bar \eta^a
  c^a\!+\!\cb^a \eta^a\!+\!ih^aM^a\!+\!\bar K_\mu^a sA_\mu^a\!+\!\bar
  L^asc^a\right\}  \nn&+\sum_{k=2}^n\int_{x,\underline{\theta}}\left\{P^0_kn^0_k+P^a_kn^a_k+\bar Q^0_ksn^0_k+\bar Q^a_ksn_k^a\right\}
\end{align}
and consider the Legendre transform of the functional $W=\ln\int\mathcal D\phi \,e ^{-S+S_1}$ with respect to the sources
$J^a_\mu$, $\eta^a$, $\bar \eta^a$, $M^a$ and $P_k^a$, where $\phi$ collectively denotes the (super)fields $A^a_\mu$, $c^a$, $\cb^a$, $h^a$ and $n_k^a$.  It is a straightforward procedure to derive the desired identities. We obtain, for the BRST symmetry  \eqn{eq_brst}-\eqn{eq_brst2},
\begin{equation}
  \label{eq_ST_BRS}
  \begin{split}
    &\int_x\left\{\frac{\delta \Gamma}{\delta A_\mu^a}\frac{\delta
      \Gamma}{\delta \bar K_\mu^a}+\frac{\delta \Gamma}{\delta
      c^a}\frac{\delta \Gamma}{\delta \bar L^a}-ih^a\frac{\delta
      \Gamma}{\delta \bar c^a}-\beta_0 c^a\frac{\delta \Gamma}{\delta
      ih^a}\right\} \\
     &+\sum_{k=2}^n\int_{x,\underline{\theta}}\left\{{1\over g^{1/2}(\underline{\theta})}P_k^0\frac{\delta
      \Gamma}{\delta \bar Q_k^0}+ {1\over g(\underline{\theta})}\frac{\delta
      \Gamma}{\delta n_k^a}\frac{\delta \Gamma}{\delta \bar
      Q_k^a}\right\}=0
    \end{split}
\end{equation}
and, for the symmetries $\delta_k^a$, (no sum over $k$)
\begin{align}
  \label{eq_ST_color}
\int_x&\left\{-  \bar Q_k^0 \frac{\delta
  \Gamma}{\delta \bar Q_k^a}+  \bar Q_k^a \frac{\delta
  \Gamma}{\delta \bar Q_k^0}+\epsilon^{abc}\left(n_k^b \frac{\delta
  \Gamma}{\delta n_k^c}+\bar Q_k^b \frac{\delta
  \Gamma}{\delta \bar Q_k^c}\right) \right.\nn
  &\,\,\,\,+\left.{g^{1/2}(\underline{\theta})}P_k^0n_k^a+{1\over g^{1/2}(\underline{\theta})}\frac{\delta
  \Gamma}{\delta P_k^0} \frac{\delta \Gamma}{\delta n_k^a}\right\}=0,
\end{align}
where the factors $g(\underline{\theta})$ account for the curved Grassmann directions.
Observe that the identities \eqn{eq_ST_color} are not integrated in $\underline{\theta}$, which follows from the fact that the symmetries $\delta^a_k$ are local in Grassmann space. 

Dimensional analysis shows that the Grassmann variables have mass dimension $-1$, which implies that $\int _{x,\underline \theta}$ has dimension $-2$ in $d=4$. This enables one to determine the dimension  of the various fields and sources, summarized below together with their ghost numbers.

\begin{table}[h!]
  \centering
  \begin{tabular}{|l|c|c|c|c|c|c|c|c|c|}
\hline
field&$A$&$c$&$\cb$&$ih$&$n$&$\bar K$&$\bar L$&$ P$&$\bar Q$\\
\hline
dim.&1&1&1&2&0&2&2&2&1\\
\hline
ghost nb. &0&1&-1&0&0&-1&-2&0&-1\\
\hline
  \end{tabular}
\label{tab_dim}
\vspace{-0.5cm}
\end{table}

The superfields $n$ have dimension $0$ in $d=4$, which makes the analysis of the scalar sector very similar to the usual nonlinear sigma model in $d=2$.
It is  straightforward, although too lengthy to be reproduced here, to prove the renormalizability of our theory. One first writes the possibly divergent (local) contribution to the effective action as $\Gamma_{\rm div}=\int_x\,{\cal L}_{\rm div}$, where ${\cal L}_{\rm div}$ contains all terms of mass dimension less or equal to four that are compatible with the linear symmetries listed previously. Further imposing the constraints \eqn{eq_ST_BRS} and \eqn{eq_ST_color}, we get, after some calculations,
\begin{equation}
  \label{eq_gamma_div}
  \begin{split}
    {\cal L}_{{\rm div}}&={\cal L}_1-\bar K_\mu^a
    s_r A_\mu^a- \bar L^a s c^a\\
    &+\sum_{k=2}^n\int_{\underline{\theta}}\left\{{\cal L}_k-P^0_kn^0_k-\bar Q^0_ksn^0_k-\bar Q^a_ksn_k^a\right\}\,,
  \end{split}
\end{equation}
with the renormalized BRST variation:
\begin{equation}
  \label{eq_brst_renorm}
    s_rA_\mu^a=Z_c^{-1} \partial_\mu c^a+g_0\epsilon^{abc}A_\mu^bc^c,
\end{equation}
and where the vanishing source terms can be written as
\begin{align}
\label{eq:ren1}
  {\cal L}_1&=\frac {1}{4Z_A}\Big(\partial_\mu A_\nu^a-\partial_\nu A_\mu^a+g_0Z_c\epsilon^{abc}A_\mu^bA_\nu^c\Big)^2+\frac {\beta_0Z_c}{2}(A_\mu^a)^2\nn
  &+\partial_\mu\cb^a\Big(Z_c^{-1}\partial_\mu c^a+g_0\epsilon^{abc}A_\mu^bc^c\Big)+ih^a\partial_\mu A_\mu^a
\end{align}
and
\begin{equation}
\label{eq:renrep}
{\cal L}_k={Z_c}\tr\left(A_\mu-\frac{i}{g_0Z_c}{\cal V}_k^{-1}\partial_\mu{\cal V}_k\right)^2\,.
\end{equation}

We see that $\Gamma_{\rm div}$ has the very same structure as the original classical action up to the two (divergent) constants $Z_A$ and $Z_c$. It is remarkable that $\Gamma_{\rm div}$ only involves two divergent constants as in the usual FP Landau gauge. Notice finally that each replica \eqn{eq:renrep} gives a gluon mass contribution $\beta_0Z_c(A_\mu^a)^2/2$ after integration over Grassmann variables. Each such term is renormalized just as the one in \eqn{eq:ren1}, as expected from the replica symmetry.

\section{Feynman rules}

Perturbation theory is most conveniently formulated in the supersymmetric formalism, which makes transparent the (dramatic) consequences of the supersymmetries -- the isometries of the curved Grassmann space -- for loop diagrams. To formulate Feynman rules, we parametrize the constrained SU($N$) superfields ${\cal V}_k$ as
\begin{equation}
\label{eq:param2}
  \mathcal V_k=\exp\left\{{ig_0\Lambda_k}\right\},
\end{equation}
where the superfields $\Lambda_k^a$ are unconstrained. 

Expanding the action \eqn{eq:action} in powers of the (super)fields to quadratic order, we obtain the various free propagators of the theory, written below in Euclidean momentum space. Note that, because of the curvature of the Grassmann subspace, it is of no use to introduce Grassmann Fourier variables. The gluon propagator reads:
 \begin{equation}
  \label{eq_propagAA}
   \left[ A^a_\mu(p)\,A^b_\nu(-p)\right]=\frac{\delta^{ab}}{p^2+n\beta_0}\left(\delta_{\mu\nu}-\frac{p_\mu
      p_\nu}{p^2}\right),
\end{equation}
where the square brackets represent averaging with the action (\ref{eq:action}).
It is transverse in momentum, as a result of the Landau gauge condition, and massive, with (bare) square mass $m_0^2=n\beta_0$, as a result of our particular gauge fixing procedure. Each replica contributes $\beta_0$ to the square mass as already clear from \Eqn{eq_action2}. The ghost propagator assumes the usual form:
\beq
   \label{eq_propagcc}
\left[ c^a(p)\,\cb^b(-p)\right]={\delta^{ab}}/{p^2}\,.
\eeq
The superfield propagator is given by
\begin{equation}
  \label{eq_propagLL}
  \left[\Lambda^a_k(p,\underline{\theta})\,\Lambda^b_l(-p,\underline{\theta}')\right]=\delta^{ab}\,\delta_{kl}\,\delta(\underline\theta-\underline\theta')/{p^2}\,,
\end{equation}
where $\delta(\underline\theta-\underline\theta')=g^{-1/2}(\underline{\theta})\,(\thetab-\thetab')(\theta-\theta')$ is the covariant Dirac delta function on the curved Grassmann space: $\int_{\underline{\theta}}\delta(\underline\theta-\underline\theta')f(\underline{\theta})=f(\underline{\theta}')$.\footnote{The correlators involving $h$ are $[h^a(p)h^b(-p)]=\delta^{ab}\beta_0/p^2$, $[h^a(p)A^b_\mu(-p)]=\delta^{ab}p_\mu/p^2$ and $[h^a(p)\Lambda^b_k(-p,\underline{\theta})]=-i\delta^{ab}/p^2$. These do not enter in loops since there is no vertex with a $h$ leg.}

The vertices of the action \eqn{eq:action} which do not involve the superfields $\Lambda_k$ are the same as for the usual FP Landau gauge. On top of the latter, there are vertices with an arbitrary number of $\Lambda_k$ legs and either zero or one gluon leg. Note that such vertices always involve the same replica. Coupling between different replicas can only arise through loop diagrams with gluons. 

The structure of the propagator \eqn{eq_propagLL} leads to a dramatic simplification of Feynman rules. Consider a loop of $\Lambda_k$ superfields with $p$ vertices insertions. It is easy to see that vertices involving $\Lambda_k$ superfields are local in Grassmann variables. Thus the loop involves the following integral over Grassmann variables:
\begin{equation}
  \label{eq_zero}
  \int_{\underline{\theta}_1,\ldots,\underline{\theta}_p}\delta(\underline\theta_1-\underline\theta_2)\cdots\delta(\underline\theta_p-\underline\theta_1)=0\,.
\end{equation}
We conclude that loops of $\Lambda_k$ superfields vanish \cite{Matsuo:1986pu}. 

This observation has two important consequences. First, the only source of dependence in the number $n$ of replicas is the bare gluon mass in \eqn{eq_propagAA}. Second, correlators or vertex functions involving only the fields $A$, $c$ and $\cb$ do not receive any contribution from replica superfields. They are thus obtained from the very same diagrams as in the FP Landau gauge, with usual YM vertices and with propagators given by \eqn{eq_propagAA} and \eqn{eq_propagcc}. These are nothing but the Feynman rules of a massive extension of the FP gauge-fixed theory, which is a particular case of the CF model. In particular, we recover the standard FP Landau gauge for $\beta_0=0$.

It is interesting to consider the above result from a slightly
different angle. Consider integrating out the superfields ${\cal V}_k$
in \eqn{eq_average2ter} to obtain an effective theory for the YM and
FP fields ($A,c,\cb,h$) which takes into account the average
\eqn{eq_average_G} over Gribov copies. \Eqn{eq_zero} above implies
that the corresponding functional integral receives no loop
contribution and is thus (perturbatively) exact at tree
level.\footnote{That this is the case can be understood by remembering
  the fact that this integral is a way to sum over the solutions $U_i$
  of the classical field equation $\partial_\mu A_\mu^U=0$ at fixed $A$, see \Eqn{eq_gribov_FP}.}
Still the result of such an integration yields a highly non trivial
effective theory: apart from a gluon mass contribution, it
contains non local contributions of arbitrary order in the gluon
field. However, it is easy to show that the corresponding vertices are
actually longitudinal in momentum space with respect to at least two
of their gluons legs. It follows that they do not contribute to
standard YM correlators involving gluon or (anti)ghost fields since
there they are to be connected to gluon propagators, which are
transverse. Again, as long as one considers standard YM correlators,
the only effect of the non trivial gauge fixing procedure considered
here is a mass term for the gluon, as in the CF model. 

\section{Renormalization group flow}

The gluon and ghost two-point vertex functions have been computed at one-loop in the CF model \cite{Tissier_10,Tissier_11} with dimensional regularization. The corresponding one-loop expressions for the present theory can thus be obtained from these papers by replacing the bare gluon mass $m_0^2\to n\beta_0$. 

A non trivial issue concerns the order in which renormalization and the limit $n\to 0$ should be performed.\footnote{A similar issue arises in the context of disordered systems with the thermodynamic limit. The common understanding is that the latter should be performed first, see e.g. \cite{young}.} 
 Due to the observation that the number $n$ of replicas only appears through the bare gluon mass, the naive prescription to take first the $n\to0$ limit reduces to the usual FP gauge fixing. This is physically unsound since we expect ghost and gluon correlators to depend on $\beta_0$. 
A possible renormalization scheme would be to introduce a renormalized parameter as $\beta_0=Z_\beta\,\beta$. But again, this leads to an unsatisfactory $n\to0$ limit. Instead, we can make use of the fact that only the combination $m_0^2=n\beta_0$ appears to implement the IR safe renormalization scheme proposed in \cite{Tissier_11}. We introduce the renormalized fields and constants $A= \sqrt{Z_A} A_r$, $c= \sqrt{Z_c} c_r$, $\bar c= \sqrt{Z_c} \bar c_r$, $g_0= Z_g g$ and $m_0^2= Z_{m^2} m^2$ and impose the following renormalization conditions
\begin{align}
\label{eq_renorm_cond}
&\Gamma_A^{(2)}(p=\mu)=m^2+\mu^2,\qquad\Gamma^{(2)}_{\bar c c}(p=\mu)=\mu^2\nn
&{\rm and}\qquad  Z_g \sqrt{Z_A} Z_c =1,\hspace{.4cm}Z_{m^2} Z_A Z_c =1,
\end{align}
where $\Gamma_A^{(2)}$ and $\Gamma^{(2)}_{\bar c c}$ denote the one-loop gluon and ghost renormalized two-point vertex functions.

The renormalization group (RG) $\beta$-functions for the coupling $g$ and the mass $m$ can be
found in \cite{Tissier_11}. Here, we discuss the general properties of the RG flow. First, we find an ultraviolet (UV) attractive fixed point at $m=0$ and $g=0$. This implies that upon removing the UV regulator (continuum limit) both the bare coupling $g_0$ and the bare mass $m_0$ vanish. It is interesting to note that this behavior is compatible with taking the limit $n\to0$ at fixed $\beta_0$ along with renormalization \cite{Dotsenko_2011}. Thus we may regard the various RG trajectories which emerge from this UV fixed point as corresponding to various choices of the gauge-fixing parameter $\beta_0$.

\begin{figure}[htbp]
  \centering
  \includegraphics[width=.8\linewidth]{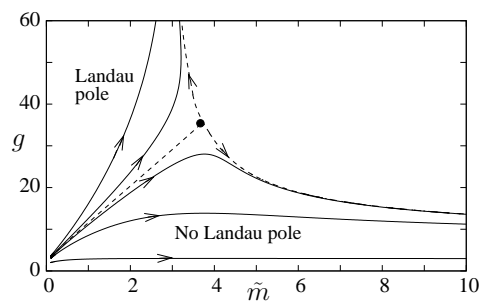}
  \caption{One-loop RG flow in the plane ($\tilde m=m/\mu,g$). The arrows indicate the flow towards the IR.}
  \label{fig_flow}
\end{figure}

For one of these trajectories $m=0$ during the whole flow, which simply corresponds to the standard FP result $\beta_0=0$. We show in Fig.~\ref{fig_flow} different trajectories obtained by
integrating numerically the one-loop $\beta$ functions.
There are two qualitatively distinct families of trajectories divided by a separatrix (dashed line on \Fig{fig_flow}). The trajectories above the latter, which include the standard FP case, are characterized by the presence of a Landau pole as one runs towards the IR: the coupling $g$ diverges at a finite momentum scale $\mu$. In constrast, for trajectories below the separatrix, the coupling constant remains bounded all the way to $\mu\to0$. These are particularly
interesting because the effective coupling constants are well-defined at all scales which means that the IR properties of
the theory may be accessible by perturbation theory. It was
actually shown in \cite{Tissier_10,Tissier_11} that one of these
trajectories leads to two-point ghost and gluon correlation functions in rather good
agreement with lattice simulations. This might seem surprising at first sight since the gauge fixings are different. However, for $\beta_0$ not too small the saddle points are suppressed in \Eqn{eq_average_G} and for $\beta_0$ not too large all minima are equiprobable. Therefore, if these conditions have some overlap for some range of $\beta_0$, we expect the usual lattice implementation of the Landau gauge to be similar to the gauge fixing proposed here.

The separatrix ends in an IR fixed point where the coupling constant is large. Its location should thus not be taken seriously. However, we expect the qualitative features of the flow described above to be stable against higher order corrections in perturbation theory. Indeed, on the one hand it is expected that the existence of a Landau
pole for the trajectory $m=0$ is a property valid at all orders and, on the other hand, trajectories lying close to the
$g=0$ axis should not be drastically influenced by higher order corrections.

\section{Conclusion}

We have devised a well-motivated, first-principle formulation of YM theories in Landau gauge, which takes into account the issue of Gribov copies in a consistent way and avoids the Neuberger zero problem. Perturbation theory shows no Landau pole and is thus potentially under control at all momentum scales. This questions the standard understanding that the low-momentum sector of YM theories is genuinely nonperturbative and opens the exciting possibility that the highly nontrivial IR physics may be accessible by perturbative methods. This opens the way to further investigations including e.g. higher-order corrections, finite temperature, or quark dynamics.

Another line of development concerns the study of classical solutions of the gauged-fixed theory proposed here. It was pointed out long ago that the classical equations of motion of a similar gauged nonlinear sigma model present non-trivial vortex solutions which may be related to confinement \cite{Cornwall:1979hz}.

Finally, it is an interesting question whether the gauge fixing procedure proposed here can be (approximately) implemented in nonperturbative continuum approaches such as Schwinger-Dyson equations or the functional RG \cite{Alkofer:2000wg} or in numerical simulations on the lattice, in the spirit of \cite{Henty:1996kv}.  This is clearly a difficult task since the average \eqn{eq_average_G} includes all extremas of the functional $S_{\rm W}[A^U]$. As a first step an average over minima is certainly feasible. More detailed studies of Gribov copies, such as in \cite{Hughes:2012hg} and, in particular of the influence of saddle points would be of great interest.

We thank G. Tarjus and N. Wschebor for fruitful interactions during the early stages of this work and for valuable suggestions concerning the limit $n\to0$ as well as L. Cugliandolo, B. Delamotte, Vi. Dotsenko and M.  M\'ezard for useful discussions.

\end{document}